\documentclass[mathpazo]{cicp}

\begin{document}
%%%%% title : short title may not be used but TITLE is required.
% \title{TITLE}
% \title[short title]{TITLE}
\title{Finite elements modelling of scattering problems for flexural waves in thin
plates: Application to elliptic invisibility cloaks, rotators and
the mirage effect}

%%%%% author(s) :
% single author:
% \author[name in running head]{AUTHOR\corrauth}
% [name in running head] is NOT OPTIONAL, it is a MUST.
% Use \corrauth to indicate the corresponding author.
% Use \email to provide email address of author.
% \footnote and \thanks are not used in the heading section.
% Another acknowlegments/support of grants, state in Acknowledgments section
% \section*{Acknowledgments}
\author[M.~Farhat]{Mohamed Farhat\corrauth, S\'ebastien Guenneau and Stefan Enoch}
%\author[S.~Guenneau]{S\'ebastien Guenneau}
%\author[S.~Enoch]{S\'ebastien Guenneau}
\address{Institut Fresnel-CNRS (UMR 6133), University of Aix-Marseille,
13397 Marseille cedex 20, France}
\email{{\tt mohamed.farhat@fresnel.fr} (M.~Farhat)}

% multiple authors:
% Note the use of \affil and \affilnum to link names and addresses.
% The author for correspondence is marked by \corrauth.
% use \emails to provide email addresses of authors
% e.g. below example has 3 authors, first author is also the corresponding
%      author, author 1 and 3 having the same address.
% \author[Zhang Z R et.~al.]{Zhengru Zhang\affil{1}\comma\corrauth,
%       Author Chan\affil{2}, and Author Zhao\affil{1}}
% \address{\affilnum{1}\ School of Mathematical Sciences,
%          Beijing Normal University,
%          Beijing 100875, P.R. China. \\
%           \affilnum{2}\ Department of Mathematics,
%           Hong Kong Baptist University, Hong Kong SAR}
% \emails{{\tt zhang@email} (Z.~Zhang), {\tt chan@email} (A.~Chan),
%          {\tt zhao@email} (A.~Zhao)}
% \footnote and \thanks are not used in the heading section.
% Another acknowlegments/support of grants, state in Acknowledgments section
% \section*{Acknowledgments}

%%%%% Begin Abstract %%%%%%%%%%%
\begin{abstract}
We propose a finite elements algorithm to solve a fourth order
partial differential equation governing the propagation of
time-harmonic bending waves in thin elastic plates. Specially
designed perfectly matched layers are implemented to deal with the
infinite extent of the plates. These are deduced from a geometric
transform in the biharmonic equation. To numerically illustrate the
power of elastodynamic transformations, we analyse the elastic
response of an elliptic invisibility cloak surrounding a clamped
obstacle in the presence of a cylindrical excitation i.e. a
concentrated point force. Elliptic cloaking for flexural waves
involves a density and an orthotropic Young's modulus which depend
on the radial and azimuthal positions, as deduced from a coordinates
transformation for circular cloaks in the spirit of Pendry et al.
[Science {\bf 312}, 1780 (2006)], but with a further stretch of a
coordinate axis. We find that a wave radiated by a concentrated
point force located a couple of wavelengths away from the cloak is
almost unperturbed in magnitude and in phase. However, when the
point force lies within the coating, it seems to radiate from a
shifted location. Finally, we emphasize the versatility of
transformation elastodynamics with the design of an elliptic cloak
which rotates the polarization of a flexural wave within its core.

\end{abstract}
%%%%% end %%%%%%%%%%%

%%%%% AMS/PACs/Keywords %%%%%%%%%%%
%\pac{}
\ams{52B10, 65D18, 68U05, 68U07} \keywords{Finite Elements;
High-Order Differential Operator; Perfectly Matched Layers;
Transformation Elastodynamics; Cloaking}

%%%% maketitle %%%%%
\maketitle

%%%% Start %%%%%%
\section{Introduction}
\label{sec1} In 2006, Pendry et al. have shown that by surrounding a
finite size object with a coating consisting of a metamaterial, we
can render this system transparent to electromagnetic radiation
\cite{pendry}. The cornerstone of this work is the geometric
transformation
\begin{eqnarray}
\left\{ \begin{array}{ll}
r^{'}=R_1+r(R_2-R_1)/R_2, & 0 < r \leq R_2,\\
\theta^{'}=\theta, & 0 < \theta \leq 2\pi,\\
z^{'}=z, & z \in \mathbb{R},
\end{array} \right.
\label{4-1}
\end{eqnarray}
where $r'$, $\theta'$ and $z'$ are radially contracted cylindrical
coordinates and $(x,y,z)$ is the Cartesian basis. This
transformation maps the disk $D_{R_2}\setminus\{0\}=\{r:\, 0<r \leq
R_2\}$ onto an annulus $D_{R_2}\setminus D_{R_1}=\{r:\, R_1 \leq r
\leq R_2\}$. In other words, if a source located in
$\mathbb{R}^2\setminus D_{R_2}$ radiates in vacuum, the
electromagnetic field cannot reach the disk $D_{R_1}$ and therefore
this region is a shelter for any object. In layman terms, a point in
the space is transformed into a disc leading to the creation of a
hole in the physical space where waves cannot penetrate from outside
but are guided around this area in a way that anything we put inside
this disc would be invisible to exterior observers, in the context
of electromagnetism \cite{pendry,pendryexp,zolla,shalaev,ieee2008},
or neutral, in the context of elastodynamics
\cite{cummernjp,sanchez,farhat08,milton,mic2009,chen07,chen}.

From a geometric point of view, the change of coordinates means that
in the annulus we should work in a stretched space with associated
metric tensor
\begin{equation}
{\bf T}={\bf J}^t_{{\bf rr'}}{\bf J}_{{\bf rr'}}/\det{({\bf J}_{{\bf rr'}})}, \label{4-2}
\end{equation}
with ${\bf J}_{{\bf rr'}}=\frac{\partial (r,\theta, z)}{\partial
(r',\theta', z')}$ the Jacobian of the transformation from free to
contracted cylindrical coordinates, ${\bf J}^{t}_{{\bf rr'}}$ its transpose
and $\det({\bf J}_{{\bf rr'}})$ its determinant.  

In terms of material parameters, the only thing to do in the annulus
is to replace the material (homogeneous and isotropic) by an
equivalent one that is inhomogeneous (its characteristics are no
longer piecewise constant but merely depend on $r'$, $\theta'$,
$z$ coordinates) and anisotropic ones (tensorial nature) whose
properties are given by \cite{nicolet,pcfbook,Ulfphil,zolla}
\begin{equation}
\underline{\underline{\varepsilon'}}=\varepsilon {\bf T}^{-1}, \quad \hbox{and} \quad
\underline{\underline{\mu'}}=\mu {\bf T}^{-1}.
\label{4-3}
\end{equation}
Variants of this formula appear in many instances in the existing
literature, such as in the book on the geometry of electromagnetism
by Post, back in 1962 \cite{post}. However, it seems that Pendry and
Ward were the first ones to apply it to simplify the computational
analysis of certain types of photonic materials in 1996 \cite{ward}.

The diagonalised form of the tensors
$\underline{\underline{\varepsilon'}}=\hbox{Diag}(\varepsilon_r,\varepsilon_{\theta},\varepsilon_3)$
and
$\underline{\underline{\mu'}}=\hbox{Diag}(\mu_r,\mu_{\theta},\mu_3)$
was given in \cite{pendry}
\begin{equation}
\varepsilon_r=\mu_r=\frac{r-R_1}{r}, \varepsilon_{\theta}=\mu_{\theta}=\frac{r}{r-R_1},
\varepsilon_3=\mu_3=\bigg(\frac{R_2}{R_2-R_1} \bigg)^2\frac{r-R_1}{r}
\label{4-4}
\end{equation}

Mimicking the heterogeneous and anisotropic nature of the tensors of
permittivity and permeability given by eq. (\ref{4-4}) became
possible only when Pendry suggested the use of newly discovered
metamaterials which are composite structures manufactured for their
exotic properties enabling one to control the electromagnetic field.
A team lead by Pendry and Smith implemented this idea using a
metamaterial consisting of concentric layers of Split Ring
Resonators (SRR), which made a copper cylinder invisible to an
incident plane wave at $8.5$ GHz as predicted by the numerical
simulations \cite{pendryexp}. Independently, Leonhardt studied
conformal invisibility by solving the Schr\" odinger equation which
is valid in the geometric optics limit \cite{leonhardt}.

A different path to invisibility was followed by McPhedran et al.
\cite{milton2} who proposed to cloak a countable set of line sources
using anomalous resonance when it lies in the close neighbourhood of
a cylindrical coating filled with a negative index material which is
nothing but a cylindrical version of the celebrated perfect lens of
Sir John Pendry \cite{pendry_prl00}.

An impedance matching route to invisibility was further developed by
Alu and Engheta \cite{engheta}. Although very promising, their
proposal relies on a specific knowledge of the material properties
of the object being concealed. Recently, these authors gave the
experimental proof of their idea \cite{alu-exp}.

Other routes to invisibility are based on homogenization; A team led
by Shalaev \cite{shalaev} has shown the possibility to make an
object nearly invisible in TE polarization. A reduced set of
material parameters was introduced to relax the constraint on the
permeability, thus leading to an impedance mismatch with vacuum
which was shown to preserve the cloak effectiveness to a good
extent.

Farhat et al. \cite{farhat} analyzed cloaking of transverse electric
(TE) fields through homogenization of radially symmetric metallic
structures. The cloak consists of concentric layers cut into a large
number of small infinitely conducting sectors which is equivalent to
a highly anisotropic permittivity. This structure was shown to work
for different wavelengths provided they are
ten times larger than the outermost sectors.\\

Soon after, Cummer and Schurig \cite{cummernjp} analyzed the
$2D$ acoustic cloaking for pressure waves in a transversely
anisotropic fluid by exploiting the analogy with TE electromagnetic
waves. Torrent and Sanchez-Dehesa \cite{sanchez} subsequently
investigated this cloaking for concentric layers of solid lattices
behaving as anisotropic fluids in the homogenization limit. Using a
similar approach, Farhat et al. \cite{farhat08} independently
demonstrated cloaking of surface liquid waves using a
micro-structured metallic cloak which was experimentally validated
at $10$ Hz. Quite remarkably, Chen and Wu \cite{chen07} and Cummer
et al. \cite{pendryprl} noticed that a $3D$ acoustic cloaking for
pressure waves in a fluid can be envisaged since the
wave equation retains its form under geometric changes.\\

However, when one moves to the domain of elastodynamics, Milton,
Briane and Willis \cite{milton} have shown that the governing
equations are not invariant under coordinate transformations and
consequently that if cloaking exists for such type of waves, it
would be of a different nature than its acoustic and electromagnetic
counterparts. Indeed, the equation of propagation of elastic waves
in solids (Navier equation) with a time harmonic dependence can be
written in the weak form:
\begin{equation}
\nabla\cdot{\bf C}:\nabla{\bf u}+\rho_0\omega^2 {\bf u}={\bf 0} \; ,
\label{navier}
\end{equation}
where ${\bf u}(x,y,z,t)={\bf u}(x,y,z)\,e^{-i\,\omega\,t}$, is the
three-component vector displacement field, $\rho_0$ is the scalar density of the
elastic medium, ${\bf C}$ is the rank $4$ elasticity tensor,
$\omega$ is the wave angular frequency,  and $t$ is the time.\\
Moreover, Milton, Briane and Willis noticed \cite{milton} that by
introducing the geometric transform ${\bf x}\to {\bf x}'$ such that
${\bf u}'({\bf x}')={\bf A}^{-T}{\bf u}({\bf x})$ with
$A_{ij}=\partial x'_i/\partial x_j$, thereby enforcing the symmetry
of the elasticity tensor, equation (\ref{navier}) takes the form:
\begin{equation}
\nabla'\cdot({\bf C'}+{\bf S'}):\nabla'{\bf
u'}+{\underline{\underline{\rho}}}'\omega^2 {\bf u'}={\bf
D'}:\nabla'{\bf u'} \; , \label{snavier}
\end{equation}
which importantly preserves the symmetry of the new elasticity
tensor ${\bf C'}+{\bf S'}$. However, this transformed equation
contains ${\bf S}'$ and ${\bf D}'$ which are two rank 3 (symmetric)
tensors such that $D'_{pqr}=S'_{qrp}$ and $\rho'_{pq}$ is a rank $2$
tensor whose expressions can be found in \cite{milton}.

Nevertheless, Brun, Guenneau and Movchan recently discovered that if
one does not constrain the symmetry of the elasticity tensor,
equation (\ref{navier}) takes the form for the case of fully coupled
in-plane shear and pressure waves:
\begin{equation}
\nabla\cdot{\bf C}':\nabla{\bf u}+\rho'\omega^2 {\bf u}={\bf 0} \; ,
\label{navier}
\end{equation}
where ${\bf C}'$ is a rank four tensor with only the main symmetries
and $\rho'$ a scalar quantity, both of them spatially varying, see
\cite{mic2009}.

Farhat, Guenneau, Enoch and Movchan further considered the case of
biharmonic bending waves propagating in thin plates and have shown
that by considering a radially dependent isotropic mass density and
a radially dependent and orthotropic flexural rigidity, it was
possible to extend cloaking by refraction to the case of
out-of-plane elastic waves \cite{prbihar}.

In this paper, we give a numerical analysis of the biharmonic
problem based on the Finite Elements Method (FEM) which is an
approximation technique for the partial differential equations (PDE)
and require a variational formulation of the problem to be studied
(expressing expressions in a weak form). Indeed, the COMSOL
formulation of the fourth order biharmonic equation is given,
supplied with appropriate boundary conditions (clamped or
stress-free) and Perfectly Matched Layers (PMLs). In the last
section, elliptic cloaking is numerically performed and the rotator and
mirage effects established for these waves confirming that they behave in a
way similar to electromagnetic and acoustic waves.

\section{Finite Element modelling for the biharmonic equation}
The equations for bending of plates are well known and can be found
in many textbooks, such as those of Timoshenko or Graff
\cite{timoshenko}-\cite{graff}. The displacement is
$W(x,y,t)$ in the $z$-direction. We choose to work in
cylindrical coordinates with a time harmonic dependence {\it i.e.}
${\bf u}(x,y,t)=(0,0,W(x,y)\,e^{-i\,p\,t})$, where $t$ is the time
and $p$ is the angular frequency.

The wavelength $\lambda$ is supposed to be large enough compared to
the thickness of the plate $h$ and small compared to its in-plane
dimensions  $h \ll \lambda \ll L$. In this case we can adopt the
hypothesis of the theory of Von-Karman.

To derive the equation of motion of the flexural out-of-plane waves,
we can use a variational point of view \cite{landau}.

\subsection{Weak form associated with the fourth-order equation}
\label{weak}

\noindent We now consider the following fourth order elliptic
equation
\begin{equation}
a\nabla\cdot({\bf b}\nabla(a\nabla\cdot({\bf b}\nabla W))) = f
\qquad \hbox{in} \quad \Omega, \label{elliptic-1}
\end{equation}
where $a$ is a scalar, and ${\bf b}$ is a matrix describing the
material of the domain, and $f$ a source term. This equation is
supplied with 'Dirichlet' (or clamped) boundary conditions $u=0$ and
${\bf n}\cdot\nabla W =0$ on $\partial \Omega$ (we note that the
second boundary condition is needed since we are in presence of a
fourth-order partial differential equation).

\noindent By multiplying Equation (\ref{elliptic-1}) with a test
function $W'$ and integrating on the domain $\Omega$
\begin{equation}
\int_{\Omega}d\tau\big[(\nabla\cdot({\bf b}\nabla(a\nabla\cdot({\bf
b}\nabla W)))W'
%+a \omega^4 u v
\big] = \int_{\Omega} d\tau f W' \, .
\label{}
\end{equation}

\noindent Integrating by parts, we obtain
\begin{equation}
-\int_{\Omega}d\tau\big[(a\nabla\cdot({\bf b}\nabla
W))(a\nabla\cdot({\bf b}\nabla W))
%+a \omega^4 u v
\big]
%-\int_{\partial\Omega}ds(-qu+g)v
= \int_{\Omega} d\tau fW ,
\label{}
\end{equation}
where we have used the fact that $W$ and its normal derivative
vanish on $\partial\Omega$.

\noindent Importantly, we need to check that this problem is well
posed. For this, we denote by $L^2(\Omega)$ (resp. ${\bf
L}^2(\Omega)$) the (class of) square integrable functions on
$\Omega$, with values in $\mathbb{C}$ (resp. $\mathbb{C}^2$). We
also introduce the Hilbert space
\begin{equation}
H^3(\Omega)=\{v\in L^2(\Omega) \; : \; \nabla v \in {\bf
L}^2(\Omega) \; , \; \Delta v \in L^2(\Omega) \; ,\; \nabla\Delta v
\in {\bf L}^2(\Omega) \}
\end{equation}

\noindent If we further assume that $W=0$ and ${\bf n}\cdot\nabla
W=0$ on $\partial\Omega$, the above space is denoted
$H^3_0(\Omega)$.

\noindent The existence and uniqueness of the solution is then
straightforwardly ensured by the Lax-Milgram theorem applied for
$f\in L^2(\Omega)$, and $(u,v)\in H^3_0(\Omega)\times
H^3_0(\Omega)$, provided that there exist $m_1$, $m_2$, $M_1$,
$M_2>0$ such that $M_1\geq a \geq m_1$ and $M_2\geq \kappa^t{\bf
b}\kappa\geq m_2$ for all $\kappa\in\mathbb{C}^2$.
%the identity
%$\int_{\partial\Omega}ds(-qu+g)v=\int_{\partial\Omega}ds {\bf
%n}\cdot(c\nabla u)v$.

\subsection{Discretisation of the variational equation}
\label{fem}

To discretise this variational problem, the basic idea is to
approximate $W$ with a linear combination of simple functions
$\phi_{i}(x,y)$ called form functions and spanning a functional
space of finite dimension (i.e. a Galerkin space with a finite
number of degrees of freedom). Then, this development is inserted
into the weak form of the fourth-order partial differential equation
thus generating a finite system of equations.

It is common to consider a triangular mesh with a reference cell
$\{x\geq 0 \; , \; y\geq 0 \; : \; x+y\leq 1\}$ the discussion we
assume a square mesh. We then consider e.g. a basis of second order
polynomials $\{1,\; , x \; , \; y \; , \; x^2 \; , y^2 \; , \; xy\}$
which spans a space of dimension 6. High-order polynomial basis can
also be used.
%Assuming that our mesh consists of $N^2$ cells:
%$I_{i,j}=[x_i,x_{i+1}[\times [y_j,y_{j+1}[$.
\noindent The function $W(x,y)$ can be written as
\begin{equation}
W=\sum_{i}^N \beta_{i}\phi_{i}(x,y) \label{dev1}
\end{equation}
where $\beta_{i}$ are values of $u$ at the node points of the
triangular mesh.
%$(x_i,y_j)$.
%Linear elements means that on each cell of the mesh the continuous function
%$u$ is bilinear.
The form functions satisfy the condition
$\phi_{i}\phi_j=\delta_{ij}$ ($\delta$ is the Kronecker symbol).

Finally using the development of $W$ (\ref{dev1}), we obtain the
system of n equations with n unknowns where we replaced $W'$ by the
set of form functions $\phi_{i}$
\begin{equation}
-\sum_{j=1}^N \bigg(
%\int_{\Omega}dxdy
\int_{\Omega}d\tau\big[(a\nabla\cdot({\bf b}\nabla
\phi_{j}))(a\nabla\cdot({\bf b}\nabla \phi_{i}))
%+a \phi_{ij}\phi_{kl}
\big] \bigg)
%-\int_{\partial\Omega}ds(-qu+g)v
=\int_{\Omega} d\tau f \phi_{i}=0 \; , \; i=1 ,..., N \; ,
%\sum_j\bigg(\int_{\Omega}dx(c\nabla\phi_j\cdot\nabla\phi_i+a\phi_j\phi_i)+\int_{\partial\Omega}ds
%q \phi_j\phi_i\bigg)U_j=
%\int{\Omega}dx f
%\phi_i+\int_{\partial\Omega} ds g \phi_i
%\, , \quad i=1...N \label{}
\end{equation}
and this can be written as a linear algebraic system
$\mathcal{K}\mathcal{U}=\mathcal{G}$ which can be solved
iteratively.

\subsection{Perfecly Matched Layers for flexural waves}
\label{pml}

\begin{figure}[h!]
\begin{center}
\scalebox{0.6}{\includegraphics[angle=0]{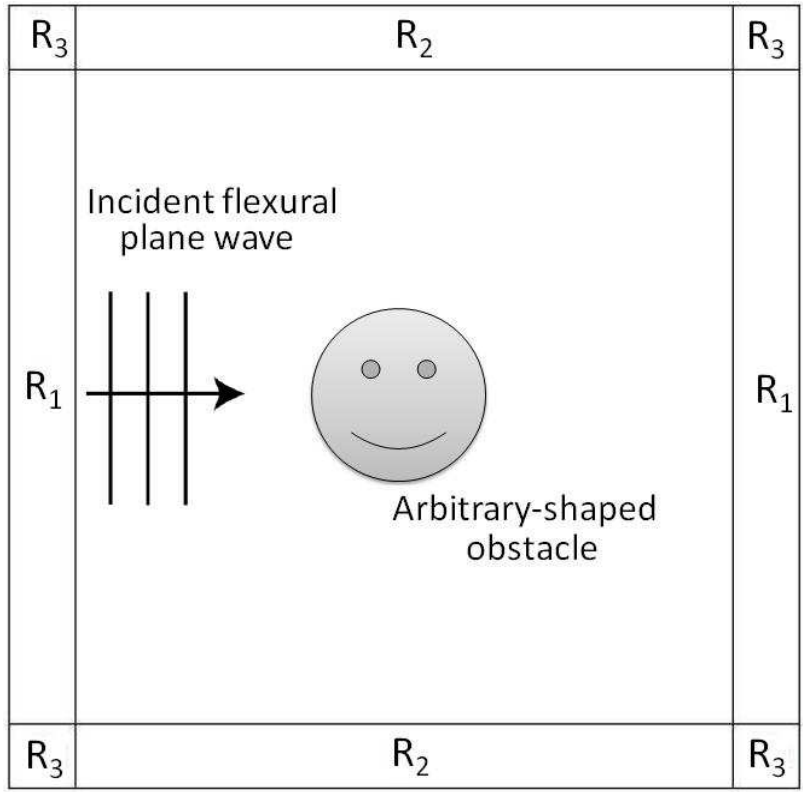}}
\caption{Schematic diagram showing the problem to be modelled: The plane elastic
wave is incident from left to right in the presence of a diffracting obstacle of arbitrary
shape. The different domains forming the PMLs regions are denoted respectively $R_1$, $R_2$ and $R_3$.}
\label{A-geometry}
\end{center}
\end{figure}

Perfectly Matched Layers (PMLs) are equivalent to a geometrical
transformation. In Cartesian coordinates they can be expressed as
\begin{equation}
x_s(x)=\int_0^x dx' s_x(x')\, \quad y_s(y)=\int_0^y dy' s_y(y')\quad
\hbox{and} \quad z_s(z)=\int_0^z dz' s_z(z'), \label{annexA-transf}
\end{equation}
whose \emph{Jacobian} is given by
\begin{equation}
{\bf J}_{x x_s}=\frac{\partial (x,y,z)}{\partial (x_s,y_s,z_s)}
=\hbox{{\bf
Diag}}\big(\frac{1}{s_x},\frac{1}{s_y},\frac{1}{s_z}\big)
\label{annexA-jacobian}
\end{equation}
where {\bf Diag} represents a diagonal matrix.

We consider now the transformation of the biharmonic equation under
an arbitrary change of coordinates (say from $(x,y,z)$ to
$(x_s,y_s,z_s)$). To do so, we have to consider the way the two
equations in the following system transform:
\begin{eqnarray}
\left \{
\begin{array}{ll}
\nabla.(-\underline{\underline{\zeta}}^{-1}\,\nabla W)+\lambda^{-1}\,V=0\\
\nabla.(-\underline{\underline{\zeta}}^{-1}\,\nabla V)+\lambda^{-1}\,\beta_0^4\,W=0
\end{array}
\right.
\label{biha-sys}
\end{eqnarray}
where $\underline{\underline{\zeta}}$ is an inhomogeneous anisotropic $2D$ tensor and
$\lambda$ is an inhomogeneous coefficient of the material 
($\underline{\underline{\zeta}}=\underline{\underline{E}}^{-1/2}$ and $\lambda=\rho^{-1/2}$).\\

For example, the first equation of this system is similar to Helmholtz's equation
except that here we have two functions $V$ and $W$ instead of a
unique function. Though we can write it in an integral form in the
domain of validity let's say $\Omega$
\begin{equation}
\int_{\Omega} d\tau\,\nabla.(-\underline{\underline{\zeta}}^{-1}\,\nabla W)
+\int_{\Omega} d\tau\,\lambda^{-1}\,V=0
\label{biha_1_int}
\end{equation}
with $d\tau$ the infinitesimal element of volume.\\
By multiplying this equation by a test function $\psi$ and integrating by parts, we obtain
\begin{equation}
-\int_{\Omega} d\tau\,\nabla\psi\cdot(-\underline{\underline{\zeta}}^{-1}\,\nabla W)+
\int_{\Omega} d\tau\,\lambda^{-1}\,\psi V=0
\label{biha_1_weak-1}
\end{equation}

Now, by operating the previously described coordinate transform
characterized by its jacobian ${\bf J}_{x_s x}={\bf J}_{x x_s}^{-1}$
that we will will note simply $J$ to lighten the notations,
(\ref{biha_1_weak-1}) transforms as
\begin{equation}
-\int_{\Omega'} d\tau' \frac{1}{\text{det}({\bf J})}\,{\bf J}^t
\nabla'\psi \cdot(-\underline{\underline{\zeta}}^{-1}\,{\bf J}^t
\nabla' W)+ \int_{\Omega'} d\tau' \frac{1}{\text{det}({\bf
J})}\,\lambda^{-1}\,\psi V=0 \label{biha_1_weak-2}
\end{equation}
where we have used the generic way the gradient and the
infinitesimal volume transform: $\nabla={\bf J}^t\nabla'$ and
$d\tau=d\tau' \frac{1}{\text{det}({\bf J})}$. Finally, using the
useful expression of the scalar product ${\bf A}\cdot{\bf B}={\bf
A}^t {\bf B}$ where ${\bf A}^t$ is the transpose of ${\bf A}$ we can
write this equation in the more adapted form
\begin{equation}
\int_{\Omega'} d\tau' (\nabla'\psi)^t \bigg\{\frac{{\bf J}
\underline{\underline{\zeta}}^{-1}\,{\bf J}^t}{\text{det}({\bf
J})}\bigg\} \nabla' W+ \int_{\Omega'} d\tau'
\bigg\{\frac{\lambda^{-1}}{\text{det}({\bf J})}\bigg\}\,\psi V=0
\label{biha_1_weak-3}
\end{equation}

\begin{figure}[!h]
\begin{center}
(a) \scalebox{0.95}{\includegraphics[angle=0]{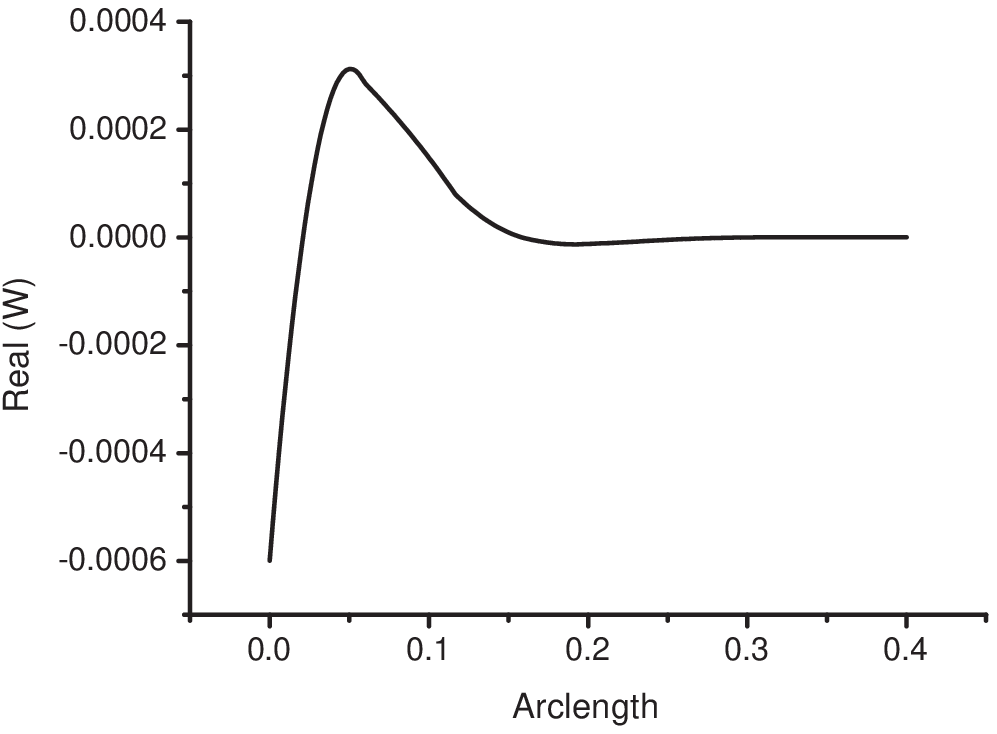}}\\

(b) \scalebox{0.95}{\includegraphics[angle=0]{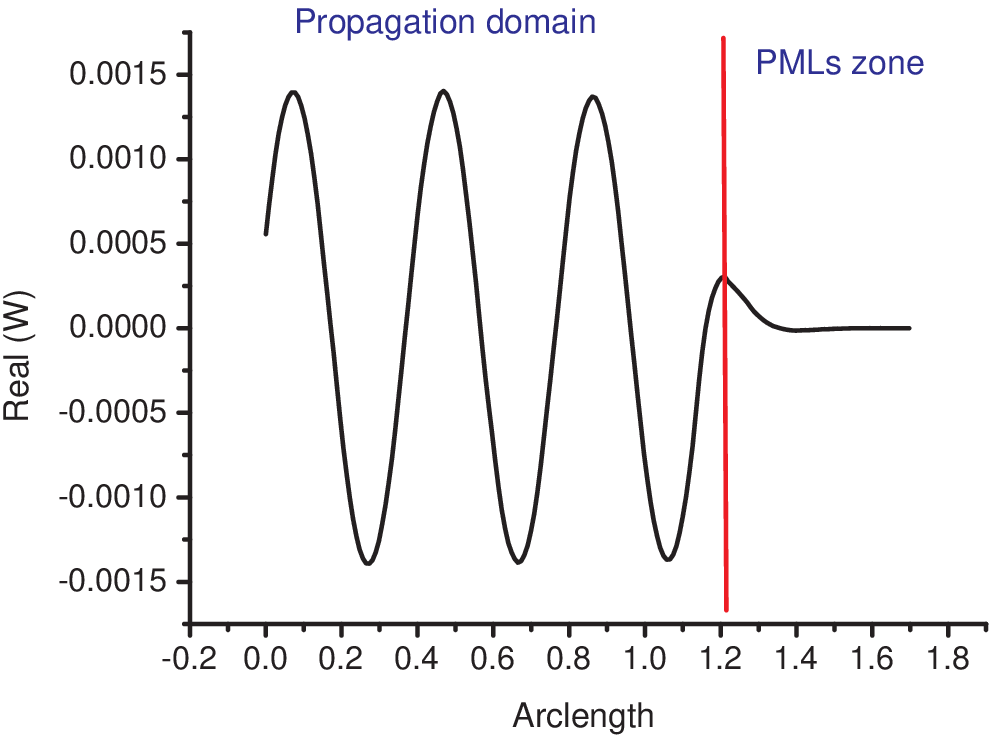}}
\caption{(a) Decay of the plane wave when it penetrates in the domain
denoted by $R_1$ or $R_2$ in Figure \ref{A-geometry} we represent here
the amplitude of the plane wave in the direction $y=0$ or $x=0$.
(b) Propagation of the plane wave along the direction $x=y$ when it lies
in the central domain (left of the red vertical line) and when it lies in the
PMLs domain $R_3$ showing clearly that its amplitude vanishes rapidly in $R_3$.}
\label{r1-r3}
\end{center}
\end{figure}

Hence, we can see from this last expression, that transforming the biharmonic equation from the initial
system of coordinates to the new one is equivalent to replacing the initial parameters
$\underline{\underline{\zeta}}^{-1}$ and $\lambda$ by the deduced parameters
$\underline{\underline{\zeta}}^{-1}$ and $\lambda '$

\begin{eqnarray}
\left \{
\begin{array}{ll}
\underline{\underline{\zeta '}}^{-1}={\bf J}_{x_s x}
\underline{\underline{\zeta}}^{-1}{\bf J}_{x_s x}^t/\text{det}({\bf J}_{x_s x})\\
\lambda '^{-1}=\lambda^{-1}/\hbox{det}({\bf J}_{x_s x})
\end{array}
\right.
\label{biha_pmls-param}
\end{eqnarray}

We can obtain the same expressions by considering the second equation in (\ref{biha-sys}).

Now, consider that we have an isotropic and homogeneous media, the use of PMLs introduce the complex matrix
$T_{PML}^{-1}$ which represents the inverse of the metric tensor. It can be expressed in terms of the parameters
of the Jacobian of the PML transformation $s_x$, $s_z$ and $s_z$

\begin{equation}
{\bf T}_{PML}^{-1}=\hbox{{\bf Diag}}(\frac{s_y s_z}{s_x},\frac{s_x
s_z}{s_y},\frac{s_x s_y}{s_z}) \label{annexA-metric}
\end{equation}

We have

\begin{eqnarray}
s_z=s_y=1\, , s_x(x)=
\left\{
\begin{array}{ll}
1\quad \hbox{In the domain of study}\\
1-i\frac{\sigma_y}{\omega \varepsilon} \quad \hbox{in the region } R_1 (\hbox{See Figure \ref{A-geometry}})
\end{array}
\right.
\label{annexA-region1}
\end{eqnarray}

For the regions $R_2$, we have just to replace $x$ by $y$ and 1 by 2 in equation (\ref{annexA-region1}), and for
the regions $R_3$, we have
\begin{equation}
s_z=1\, , \quad s_y(y)=1-i\frac{\sigma_y}{\omega \varepsilon}\, \quad \hbox{and} s_x(x)=1-i\frac{\sigma_x}{\omega \varepsilon}
\label{}
\end{equation}

Figure \ref{r1-r3} clearly confirms the mechanism of PMLs based on the transformation
of biharmonic equation from the initial system ($x,y$) to ($x_s,y_s$) representing absorbing
layers in such a way incident waves are perfectly transmitted (without any reflexions)
independently from their frequency or incidence angles.\\
To our knowledge, this is the first time PMLs were applied to the fouth order biharmonic problem.\\
For exapmle, Figure \ref{r1-r3} (b) shows the decay of an incident plane wave in the PML zone (rectangle
$R_3$ of Figure \ref{A-geometry}): the amplitude of the wave can be considered zero after only a short
propagation distance in the absorber medium.

\section{Elliptic invisibility cloaks and the mirage effect for biharmonic problems}

\subsection{Elliptic invisibility cloak for flexural waves}

\begin{figure}[!h]
\begin{center}
(a) \scalebox{0.425}{\includegraphics[angle=0]{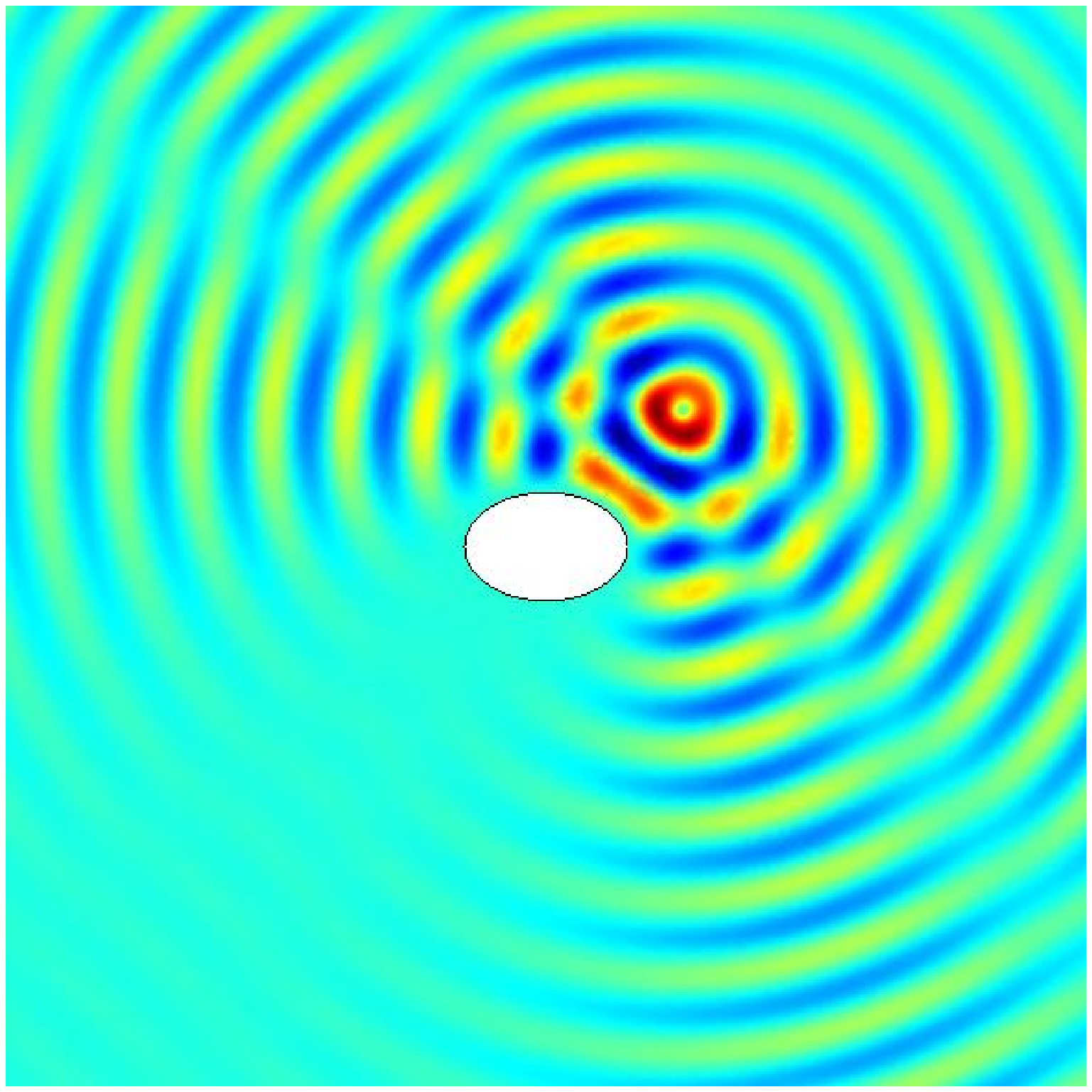}}
\hspace{0.25cm}
(b) \scalebox{0.425}{\includegraphics[angle=0]{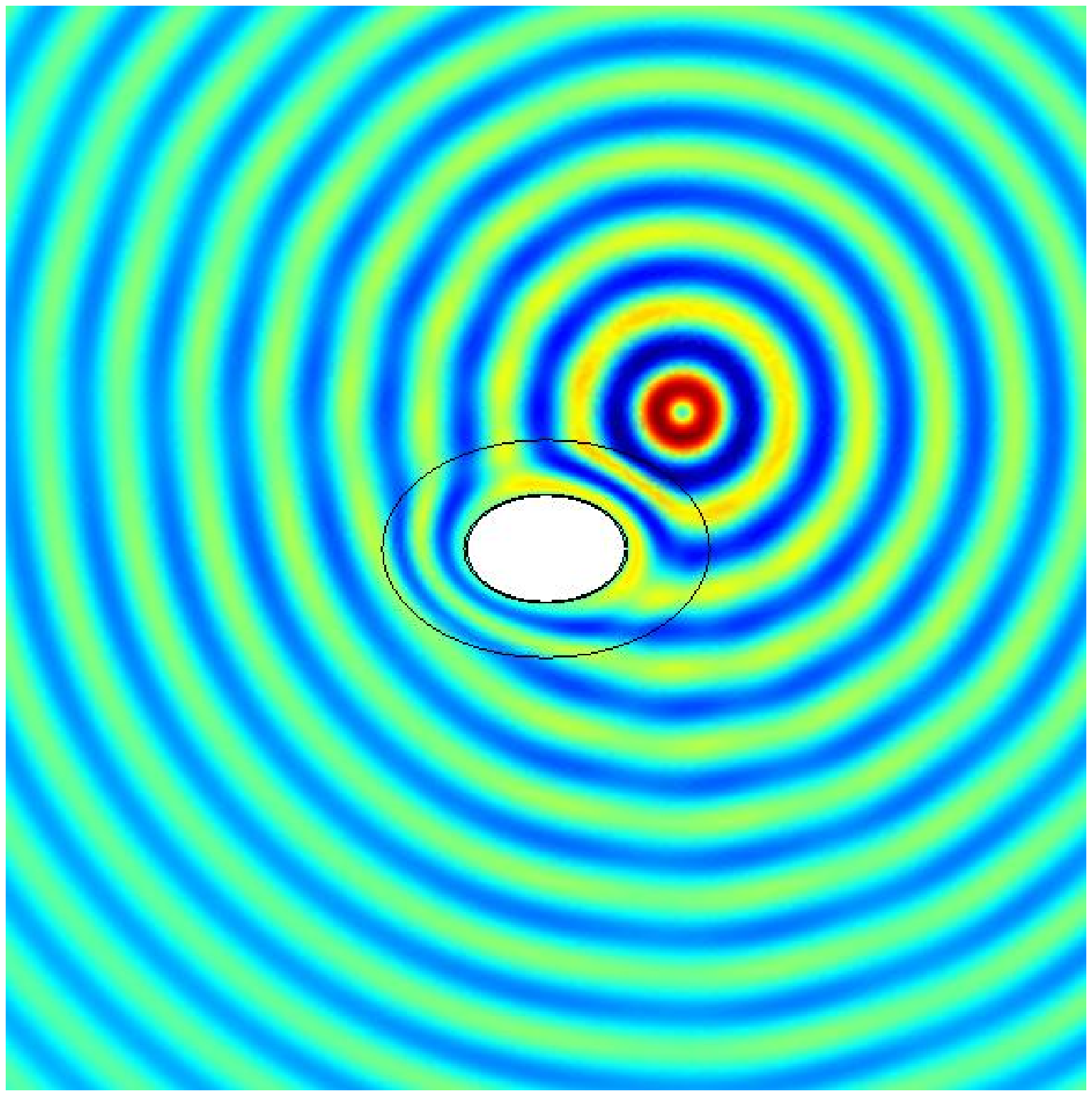}}
\caption{Right: Real part of the displacement field $W$ distribution in the vicinity of the
cloaked elliptic rigid clamped obstacle. The source is located at point
$(0.5, 0.5)$ and its wavelength is $\lambda_0=2\pi/\beta_0=0.28$,
which is of the same order as the inner and outer radii (major and minor) of the
cloak: $a_1=0.2$, $a_2=0.3$ and $b_1=0.4$, $b_2=0.6$.
Left: Real part of the displacement field $W$ scattered by a rigid clamped obstacle of
minor radius $0.2$ m and major radius $0.3$ m for an incoming cylindrical wave of the same
frequency.}

\label{elliptic_clamped}
\end{center}
\end{figure}

This may be deduced from the circular case by scaling the Cartesian
coordinates. The global transformation is a mapping of a holey
elliptic domain (the inner and outer boundaries are concentric
ellipses with the same eccentricity) on a simply connected elliptic
domain bounded by the outer ellipse of the cloak. The detailed
computation of the equivalent material properties is given by the
following sequence of transformations. The starting point is the
ellipse that bounds the exterior limit of our cloak with its
principal axes chosen conveniently parallel to the coordinates axes.
The first step is aimed at restoring the previous situation namely a
circular cloak. For this purpose, the plane is scaled by a factor
$s_y$ along the (arbitrary chosen) -axis so that the initial ellipse
becomes a circle (in the scaled coordinates) defined by the
transformation $y=s_y y_s$ (and simply $x=x_s$) characterized by the
Jacobian matrix ${\bf J}_{xx_s}={\rm {\bf Diag}}(1,s_y,1)$. The next
three transformations are then the ones used to build the circular
cloak: transformation to cylindrical coordinates, radial contraction
(the active part), and transformation to rectangular coordinates. It
is important to note that it is the scaled variable $y_s$ that is
involved in these various operations. The last step to be performed
is an inverse scaling along $y$-axis: $y'=(1/s_y)y'_s$ to recuperate
the initial elliptical shape of the cloak and an identity for the
transformation of the outside of the cloak. At the end, a cloak is
obtained whose inner and outer boundaries are ellipses with
$R_1=x$-axis of the hole, $s_y$, $R_1=y$-axis of the hole,
$R_2=x$-axis of the external boundary, $s_y$, $R_2=y$-axis of the
external boundary. The total Jacobian of this sequence of
transformations is
\begin{equation}
{\bf J}_{xx'}={\bf J}_{xx_s}{\bf J}_{x_sr}{\bf J}_{rr'}{\bf
J}_{r'x'_s}{\bf J}_{x'_sx'}
\label{invis_elliptic}
\end{equation}

The inverse of the matrix is given explicitly by \cite{ieee2008}
\begin{equation}
\begin{array}{ll}
{\bf T}^{-1}= &{\rm {\bf Diag}}(1,s_y,1)R(\theta'_s){\rm {\bf
Diag}}(\alpha,\frac{r'_s}{r},1)R(-\theta_s) {\rm
{\bf Diag}}(1,\frac{1}{s_y^2},1)R(\theta_s) \\
&{\rm {\bf Diag}}(\alpha,\frac{r'_s}{r},1)R(-\theta'_s){\rm {\bf
Diag}}(1,s_y,1) \frac{r}{\alpha r'_s} \; ,
\end{array}
\label{T-1}
\end{equation}
with $r'_s=\sqrt{x'^2+{(y'/s_y)}^2}$, $\theta_s=\theta'_s= 2
\rm{arctan}((y'/s_y)(x'+r'_s))$ and $r=(r'_s-R_1)/\alpha$.

Note the angles and distances are computed in the scaled coordinate
systems $(x_s,y_s)$ where the ellipses are mapped to circles.

Figure \ref{elliptic_clamped} (b) shows the calculated displacement field distribution when an
elastic point source vibrating in the out-of-plane direction is placed near the optimized
cloak. It is shown that in the near region of the cloak, the displacement field stays almost
unperturbed, thus a so called perfect elliptic cloak is obtained. In order to see how much
improvement has been achieved through the presence of the elliptic coat, snapshots of the
displacement $W$ in the presence of a clamped obstacle ($W=0$ and $\partial W/\partial r=0$
where $r$ describes the boundary of the ellipse) are shown in (a).

\subsection{Transformation elastodynamics: Elliptic rotator and mirage effect}

\begin{figure}[!h]
\begin{center}
\scalebox{0.7}{\includegraphics[angle=0]{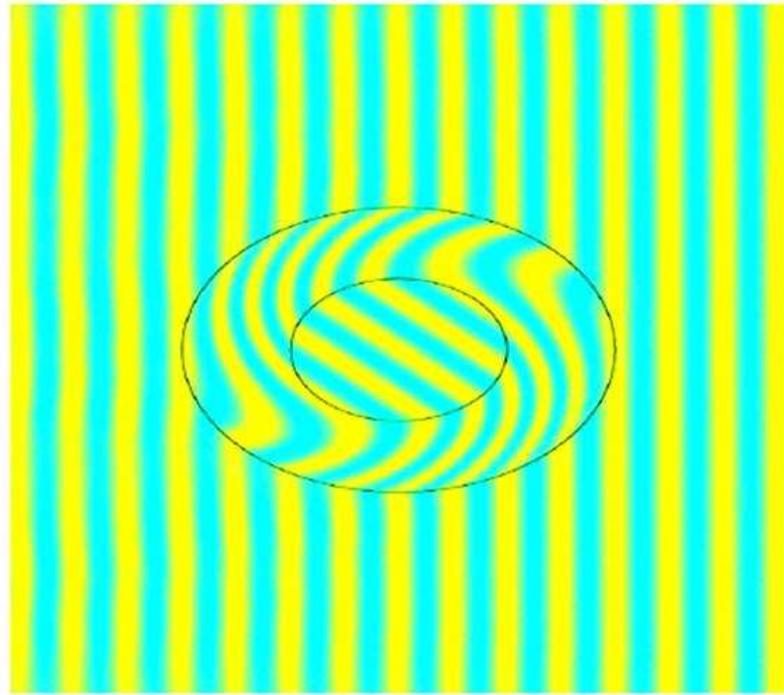}}
\caption{A plane bending wave is incident from left to right onto a
coating consisting of two concentric ellipses of the same dimensions
as in Figure \ref{elliptic_clamped}. The Young modulus and density
of the shell are deduced from the transformation given in Equation
(\ref{rotator_transfo}). We can see that the wave fronts are rotated
with an angle $\theta_0=\pi/4$ radians.} \label{rotator_fig}
\end{center}
\end{figure}

Based on the coordinate transformation methods \cite{leonhardt}, we propose here a two-dimensional
transformation-media coat that can rotate fields but remains itself invisible.
The Young modulus tensor and density of such a "field rotator" for flexural waves can
be deduced from appropriate coordinates transforms.

The difference with invisibility (considered in the previous section and where we map a
point on a circle of radius $R_1$ ) is that here, we map a circle (of radius $R_1$ or $R_2$)
onto itself. So let's consider the following transformation:

\begin{equation}
\left \{
\begin{array}{lll}
r'=r\\
\theta'=\theta+\alpha_1 r+\beta_1\\
z'=z
\end{array}
\right.
\label{rotator_transfo}
\end{equation}
with $\alpha_1=\theta_0/(R_1-R_2)$ and $\beta_1=\theta_0/(R_2-R_1)$. $\theta_0$ is the angle
of rotation of the wave fronts.\\
The \emph{Jacobian} of this transformation is given by

\begin{equation}
J_{rr'}=\frac{\partial(r,\theta,z)}{\partial(r',\theta',z')}=
\left(
\begin{array}{ccc}
1&0&0\\
-\alpha_1&1&0\\
0&0&1
\end{array}
\right)
\label{jacob_rotator}
\end{equation}
This expression may be injected in Equation (\ref{invis_elliptic}) instead of
the invisibility \emph{Jacobian} $J_{rr'}$ to obtain the matrix $T$ which gives
us the physical parameters of the coat (Young's modulus and density). In addition,
inside the inner circle of radius $R_1$, we have the additional transform
($r'=r$, $\theta'=\theta+\theta_0$ and $z'=z$) whose \emph{Jacobian} is the
$3\times3$ identity matrix and doesn't contribute thus to the materials specifications
inside the cloak. Finally, the FEM computations using these expression show in Figure
\ref{rotator_fig} the displacement filed in the vicinity of an ideal field rotator
designed by transformation-media theory.\\

We consider also another application of coordinates transformation and which
has been verified for others types of waves (acoustic and elastodynamics).

\begin{figure}[!h]
\begin{center}
(a) \scalebox{0.425}{\includegraphics[angle=0]{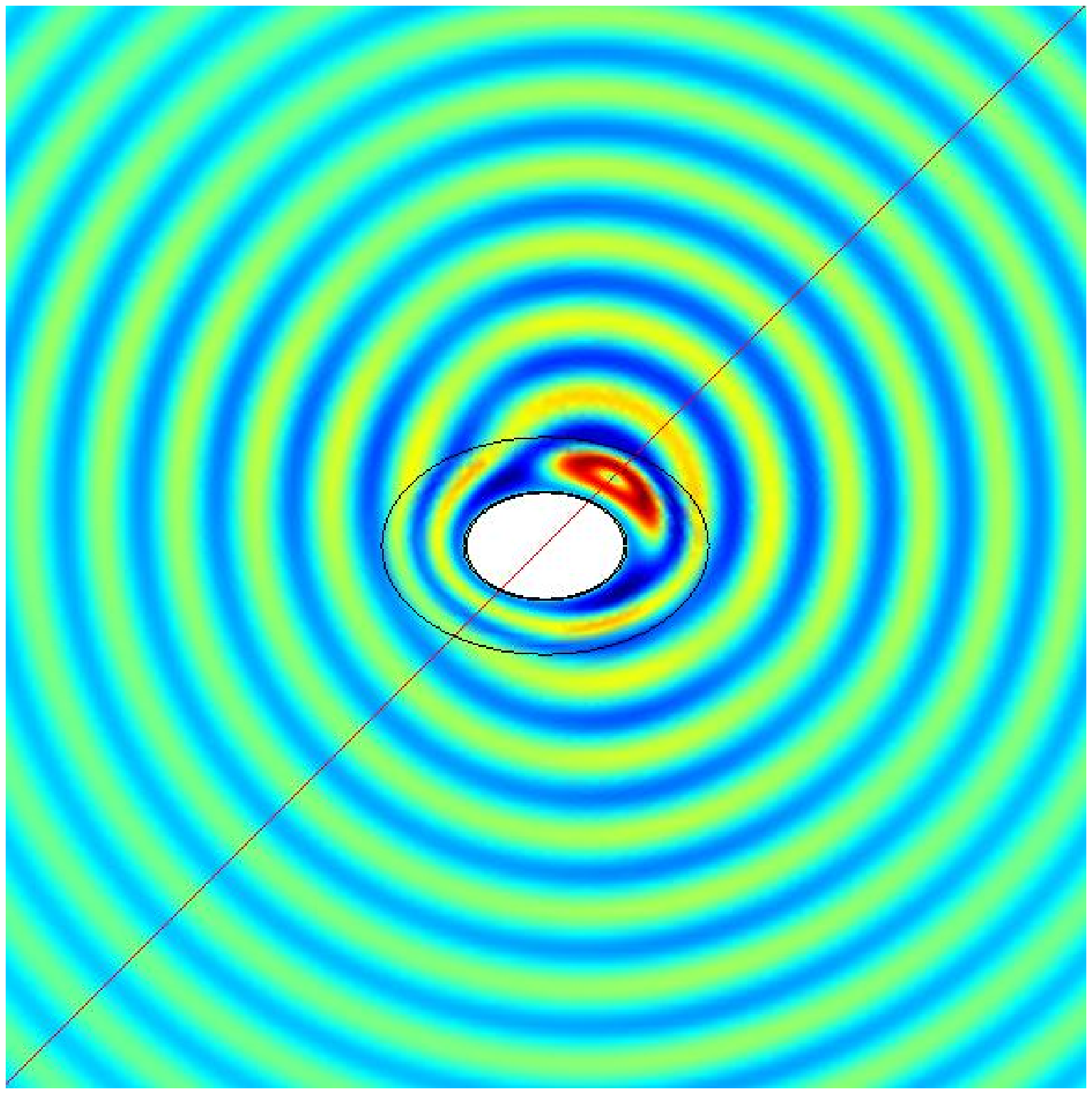}}
\hspace{0.25cm}
(b) \scalebox{0.425}{\includegraphics[angle=0]{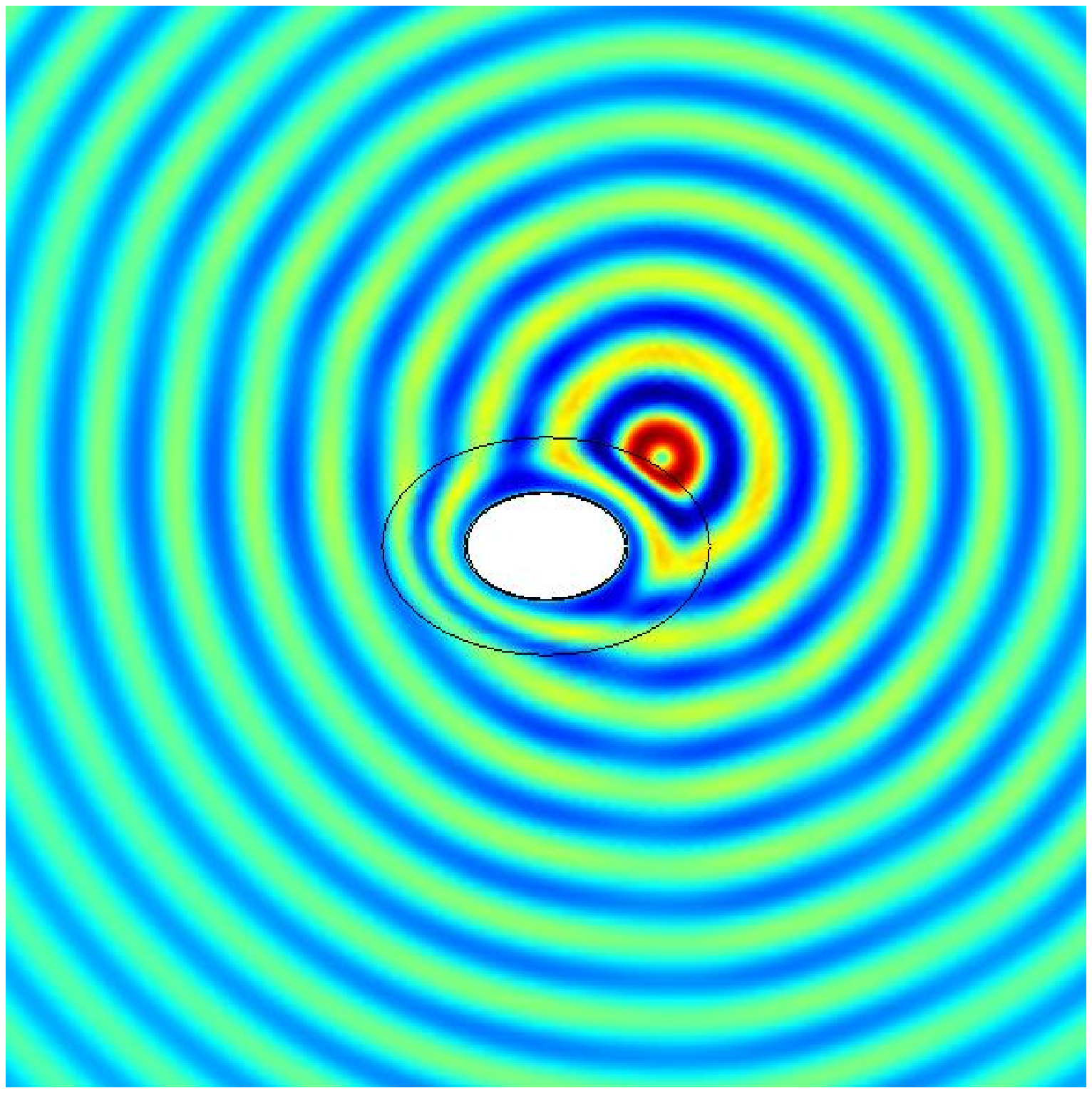}}\\
\vspace{1cm}
(c) \scalebox{1.1}{\includegraphics[angle=0]{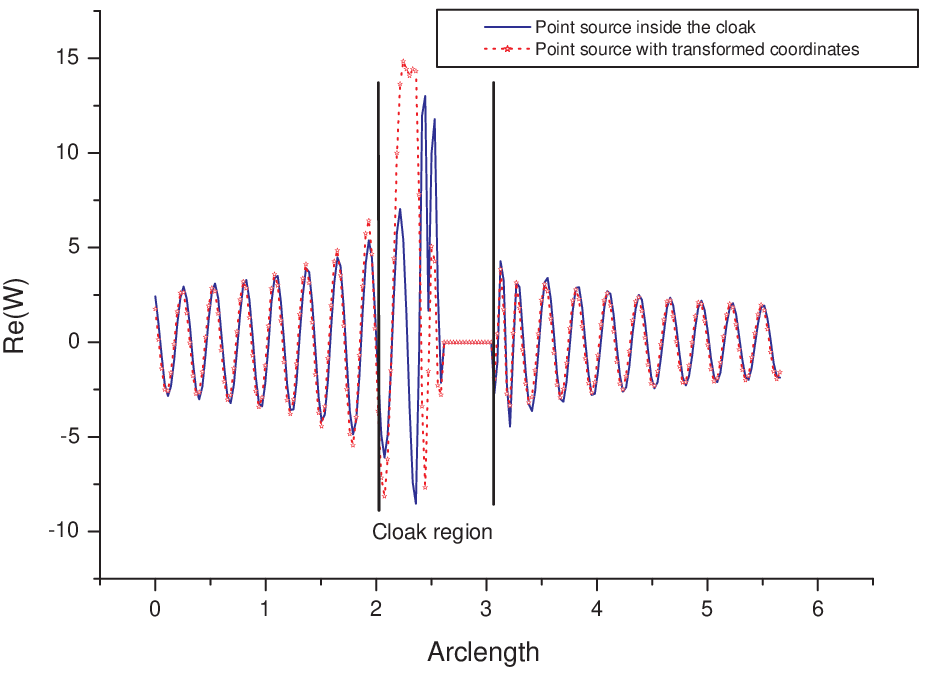}}
\caption{(a) Real part of the displacement field $W$ when the source
is located inside the coating at the point $(0.25,0.25)$. The field seems
to be emitted by a shifted source located at point $(0.425,0.325)$ (b);
(c) Real part of the displacement $W$ along the line $x=y$ for
a source located inside the coating at point $(0.25,0.25)$ as shown in
Figure \ref{mirage} (a) (solid line) and for a source located in the homogeneous
plate at point $(0.425,0.325)$ (dashed-starred line) deduced from the geometrical
transform given in (\ref{T-1}).}

\label{mirage}
\end{center}
\end{figure}

In Figure \ref{mirage} the vertical displacement $W$ outside the cloak
seems to originate from a location $r'_s$, which is slightly
shifted with respect to the real position $r_s$ of the source (located
inside the cloak itself). This effect is similar to the mirage
effect already observed in electromagnetic cloaks \cite{zolla}. The
vertical displacement on the line $x=y$ is shown for both a source
located inside the coating and a source shifted laterally at $r_s$
in a homogeneous plate (i.e., without any cloak and inclusion). The
respective positions are given in the figure caption.\\

In conclusion, we have studied in this paper some properties of the biharmonic equation
governing the propagation of flexural waves in thin elastic plates. we have shown the way
to generalize some concepts like cloaking or the rotator and the mirage effect to the domain
of elasticity in the special case of elliptic geometries. Despite the fact that general elastic
equations are not invariant under geometrical changes, we have demonstrated that cloaking
through coordinates changes is applicable to some special configurations (thin plates). Moreover
Perfectly Matched Layers have been performed using the finite elements technique for the biharmonic
problems.

%%%% Acknowledgments %%%%%%%%
\section*{Acknowledgments}
The authors would like to thank insightful discussions with Prof.
A.B. Movchan and N.V. Movchan at Liverpool University, Dr. C.G.
Poulton at the University of Technology of Sydney (UTS) and Prof.
R.C. McPhedran at the University of Sydney.

%%%% Bibliography  %%%%%%%%%%


\begin{thebibliography}{99}
\bibitem{pendry} J.B. Pendry, D. Schurig, and D.R. Smith, Science {\bf 312}, 1780 (2006).
\bibitem{pendryexp} D. Schurig, J.J. Mock, B.J. Justice, S.A. Cummer, J.B. Pendry,
A.F. Starr, D.R. Smith, Science {\bf 314}, 977 (2006).
\bibitem{zolla} F. Zolla, S. Guenneau, A. Nicolet, and J.B. Pendry, Opt. Lett. {\bf 32}, 1069-1071 (2007).
\bibitem{shalaev}
W. Cai, U.K. Chettiar, A.V. Kildiev and V.M. Shalaev,
%``Optical Cloaking with metamaterials'',
Nature {\bf 1}, 224-227 (2007).
\bibitem{ieee2008}
A. Nicolet, F. Zolla and S. Guenneau,
%``Finite-Element Analysis of Cylindrical Invisibility Cloaks of Elliptical Cross Section,''
IEEE Trans. MAg. {\bf 44}, 1150-1153 (2008)
\bibitem{cummernjp} S.A. Cummer and D. Schurig, New J. Phys. {\bf 9}, 45 (2007).
\bibitem{sanchez}
D. Torrent and J. Sanchez-Dehesa, New J. Phys. {\bf 10}, 063015
(2008).
\bibitem{farhat08}
M. Farhat, S. Enoch, S. Guenneau and A.B. Movchan, Phys. Rev. Lett.
{\bf 101}, 134501 (2008).
\bibitem{chen07} H. Chen and C. T. Chan,
Appl. Phys. Lett. {\bf 91}, 183518 (2007).
\bibitem{milton} G.W. Milton, M. Briane, and J.R. Willis, New J. Phys. {\bf 8}, 248 (2006).
\bibitem{mic2009}
M. Brun, S. Guenneau and A.B. Movchan, Appl. Phys. Lett. {\bf 94}
061903 (2009).
\bibitem{chen}
H. Chen, B.I. Wu, B. Zhang, and J.A. Kong, Phys. Rev. Lett. {\bf
99}, 063903 (2007).
\bibitem{Ulfphil} U. Leonhardt and T. G. Philbin,
%``General relativity in electrical engineering,''
New J. Phys. {\bf 8}, 247 (2006).
\bibitem{nicolet} A. Nicolet, J.F. Remacle, B. Meys, A. Genon and W. Legros,
%``Transformation methods in computational electromagnetism,'' J.
Appl. Phys. {\bf 75}, 6036-6038 (1994).
\bibitem{pcfbook} F. Zolla, G. Renversez, A. Nicolet, B. Kuhlmey, S.
Guenneau and D. Felbacq, Foundations of photonic crystal
fibres (Imperial College Press, London, 2005).
\bibitem{post}
E.G. Post, Formal Structure of Electromagnetics; General Covariance
and Electromagnetics (Interscience, 1962).
\bibitem{ward}
A.J. Ward and J.B. Pendry,
%``Refraction and geometry in Maxwell's equations,''
J. Mod. Opt. {\bf 43}, 773-793 (1996).
\bibitem{leonhardt} U. Leonhardt,
%``Optical conformal mapping,''
Science {\bf 312} 1777-1780 (2006).
\bibitem{milton2}
N.A. Nicorovici, R.C. McPhedran and G.W. Milton,
%``Optical and dielectric properties of partially resonant composites,''
Phys. Rev. B {\bf 49}, 8479-8482 (1994).
\bibitem{pendry_prl00} J.B. Pendry,
%Negative refraction makes a perfect lens,
Phys. Rev. Lett. {\bf 86}, 3966-3969 (2000).
%\bibitem{pendry_jpc03}
%J.B. Pendry and S.A. Ramakrishna, ``Focussing light using negative
%refraction,'' J. Phys. Cond. Matter {\bf 15}, 6345 (2003).
\bibitem{engheta} A. Alu and N. Engheta,
%``Achieving Transparency with Plasmonic and Metamaterial Coatings,''
Phys. Rev. E {\bf 95} 016623 (2005).
\bibitem{alu-exp} B. Edwards, A. Alu, M. G. Silveirinha, and N. Engheta,
Phys. Rev. Lett. {\bf 103} 153901 (2009).
\bibitem{farhat}
M. Farhat, S. Guenneau, A.B. Movchan and S. Enoch,
%``Achieving invisibility over a finite range of frequencies,''
Opt. Express {\bf 16}, 5656-5661 (2008)
\bibitem{pendryprl} S.A. Cummer, B.I. Popa, D. Schurig, D.R. Smith, J. Pendry,
M. Rahm, and A. Starr, Phys. Rev. Lett. {\bf 100}, 024301 (2008).
\bibitem{prbihar}
M. Farhat, S. Enoch, S. Guenneau and A.B. Movchan, Phys. Rev. B {\bf
79} 033102 (2009).
\bibitem{timoshenko} S. Timoshenko, Theory of plates and shells (McGraw-Hill, New York, 1940).
\bibitem{graff} K.F. Graff, Wave motion in elastic solids (Dover, New York, 1975).
%\bibitem{miltonbbook} G.W. Milton, {\it Theory of composites}
%(Cambridge University Press, Cambridge, 2002).
%\bibitem{kohn84}
%R.V. Kohn and M. Vogelius, ``Identification of an unknown
%conductivity by means of measurements at the boundary,'' Inverse
%Problems D. McLaughin ed., SIAM-AMS Proc. {\bf 14} 113-123 (1984)
%\bibitem{greenleaf}
%A. Greenleaf, M. Lassas and G. Uhlmann, ``On nonuniqueness for
%Calder´on's inverse problem,''Math. Res. Lett. {\bf 10}, 685-693
%(2003).
%\bibitem{norris} A.N. Norris and C. Vemula, J. Sound Vib. {\bf 181}, 115-125 (1995).
%\bibitem{evans} D.V. Evans, and R. Porter,  J. Eng. Math. {\bf 58}, 317-337 (2007).
%\bibitem{movchan} A.B. Movchan, N.V. Movchan, and R.C. McPhedran, Proc. R. Soc. A {\bf 463},
%2505-2518 (2007).
\bibitem{landau} L.D. Landau, and E.M. Lifschitz, Elasticity theory (Pergamon Press, 1954).
\end{thebibliography}
\end{document}